\documentclass[review, number, sort&compress]{elsarticle}
\biboptions{sort&compress}
\usepackage{geometry} 
\usepackage{natbib}
\usepackage{graphics}
\usepackage{threeparttable}
\usepackage{longtable}
\usepackage{lscape}
\usepackage{appendix}
\usepackage{amsmath}
\usepackage{multirow}
\usepackage{rotating}
\usepackage{epstopdf}
\usepackage{subfig}
\journal{Nuclear Instruments and Methods in Physics Research A	}

\begin{document}

\begin{frontmatter}

\title{The Track Imaging Cerenkov Experiment}

\author[efi,kavli]{S. ~A. ~Wissel\corref{cor1}\fnref{fn1}}
\ead{wissels@uchicago.edu}

\author[anl]{K. ~Byrum}

\author[loyola]{J. ~D. ~Cunningham}

\author[anl]{G. ~Drake}

\author[efi,anl]{E. ~Hays\fnref{fn2}}

\author[anl]{D.~Horan\fnref{fn3}}

\author[utah]{D. ~Kieda}

\author[anl]{E.~Kovacs}

\author[anl]{S.~Magill}

\author[anl]{L.~Nodulman}

\author[efi,kavli]{S. ~P. ~Swordy\fnref{fn4}}

\author[anl]{R. ~Wagner}

\author[efi,kavli]{S. ~P.~Wakely}

\cortext[cor1]{Corresponding author}
\fntext[fn1]{Currently at the Princeton Plasma Physics Laboratory, P. O. Box 451, Princeton, NJ 08543-0451, USA}
\fntext[fn2]{Currently at the N.A.S.A./Goddard Space-Flight Center, Code 661, Greenbelt, MD 20771, USA}
\fntext[fn3]{Currently at the Laboratoire Leprince-Ringuet, Ecole Polytechnique, CNRS/IN2P3, F- 91128 Palaiseau, France}
\fntext[fn4]{Deceased}

\address[efi]{Enrico Fermi Institute, University of Chicago, Chicago, IL 60637, USA}
\address[kavli]{Kavli Institute for Cosmological Physics, University of Chicago, Chicago, IL 60637, USA}
\address[anl]{Argonne National Laboratory, 9700 S. Cass Ave., Argonne, IL 60439, USA}
\address[loyola]{Loyola University Chicago, 1032 W. Sheridan Rd., Chicago, IL 60660, USA}
\address[utah]{Department of Physics, University of Utah, Salt Lake City, UT 84112, USA}

\begin{abstract}
We describe  a dedicated cosmic-ray telescope that explores a new method for detecting Cerenkov radiation from high-energy primary cosmic rays and the large particle air shower they induce upon entering the atmosphere. Using a  camera comprising 16 multi-anode photomultiplier tubes for a total of 256 pixels, the Track Imaging Cerenkov Experiment (TrICE) resolves substructures in particle air showers with $0.086^{\circ}$ resolution. Cerenkov radiation is imaged using a novel two-part optical system in which a Fresnel lens provides a wide-field optical trigger and a mirror system collects delayed light with four times the magnification. TrICE records well-resolved cosmic-ray air showers at rates ranging between 0.01-0.1 Hz.
\end{abstract}


\end{frontmatter}
\section{Motivation}
High-precision cosmic-ray composition measurements carried out by balloon- and satellite-borne experiments can achieve charge resolutions of less than 5\% \citep{2009ApJ...707..593A,2008ApJ...678..262A} and in some cases, can measure the isotopic abundance of these particles \citep{0004-637X-698-2-1666}. Such experiments become limited by statistics at the few TeV/amu regime, because of their small geometric factors. On the other hand, ground-based arrays can measure the composition of cosmic rays from the few TeV through EeV regime by employing the Earth's atmosphere as a calorimeter and measuring properties of the extensive air shower (EAS) produced when a high-energy cosmic ray interacts with nuclei in the atmosphere \cite{2009NuPhS.190..240T,Antoni20051}. Such indirect composition measurements rely on hadronic interaction models and have limited charge resolution. A new technique proposed by  \citep{Kieda:2000ky} and implemented by \citep{2007PhRvD..75d2004A,2010PhDT........37W} shows that by measuring the Cerenkov light initiated both directly by the cosmic ray (DC light) and by the secondary particles in the air shower (the EAS light), the composition of cosmic rays at TeV-PeV energies can be measured with $<20\%$ charge and energy resolution.  Note that while the field of view of typical satellite- and balloon-borne experiments is typically greater than imaging Cerenkov telescopes, the ground-based arrays can also expand their exposures by collecting data for longer periods of time.

\begin{figure}[htbp] 
   \centering
	\includegraphics[width=90mm]{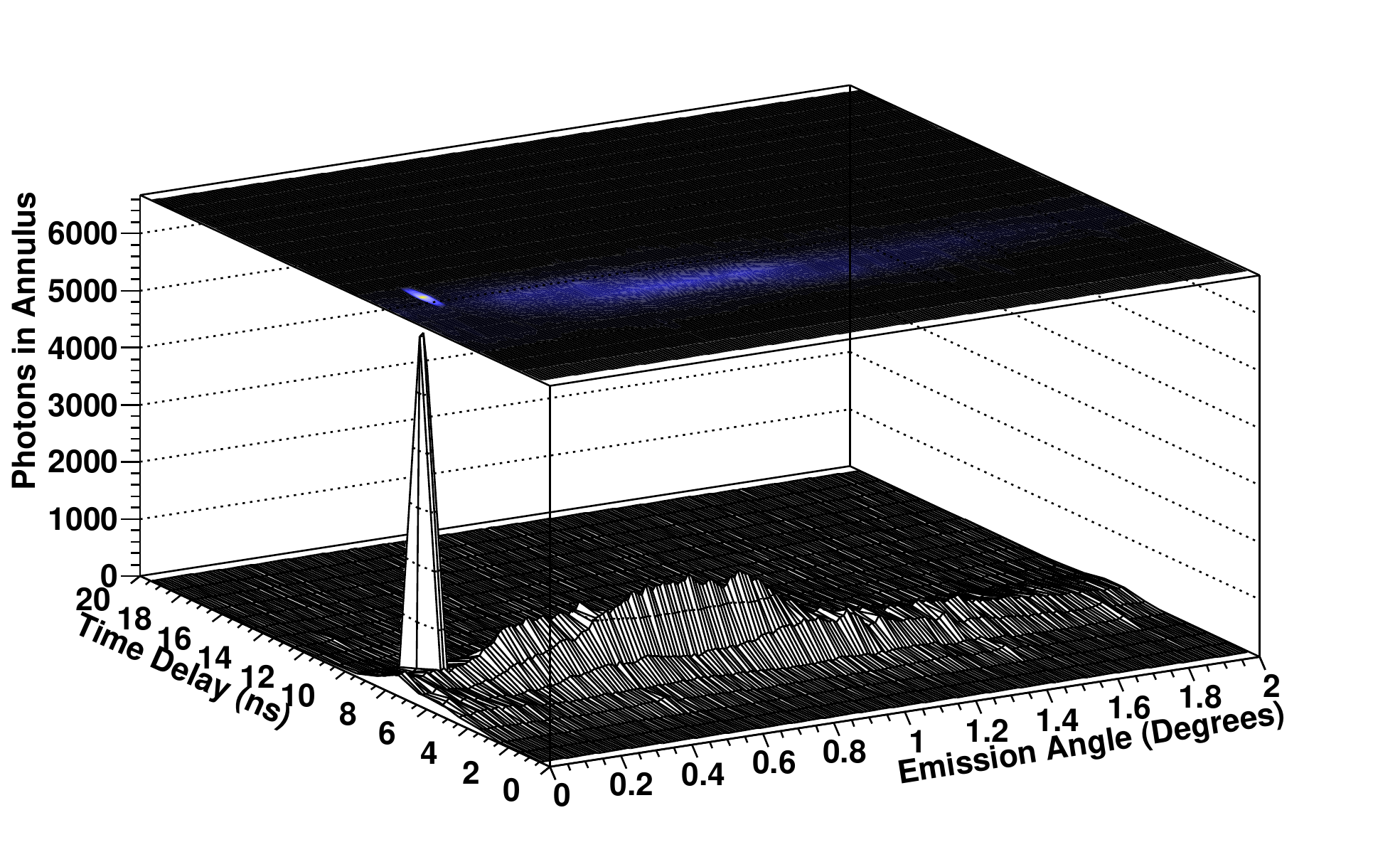}
   \caption[The timing and angular separation between DC light and EAS light]{Simulated photons emitted via Cerenkov radiation starting from the top of the atmosphere collected in a ring between 50 m and 115 m from the shower impact.. The DC component is cleanly separated from the EAS light when the following quantities are considered: the angle between the cosmic ray's trajectory and the arrival direction of the photon and the time delay from when the cosmic ray would have impacted the ground and the arrival time of the photon on the ground. The DC component is compact in both time and angular space ($0.05^{\circ}$ in angular extent and 1 ns in duration) and cleanly separated from the extensive air shower component by 2 ns and $0.2^{\circ}$ in a simulated 50 TeV $^{56}$Fe air shower that is vertically incident and interacts 25 km above ground-level.  }
   \label{fig:dc_timeangle}
\end{figure}

High-energy cosmic rays continuously produce Cerenkov radiation in the atmosphere until they interact hadronically with nuclei in the atmosphere. Thereafter, the cosmic ray generates an extensive air shower (EAS) producing $>10^{5}$ particles per shower, largely depending on the primary's energy. The light directly produced by a heavy nucleus is appreciable ($ > 3 \times 10^{4}$ for the simulation in Fig. \ref{fig:dc_timeangle}), but the difficulty lies in separating it from that produced in the extensive air shower. Above the first interaction height, the particle generates DC light with a narrow Cerenkov angle ($\sim0.1^{\circ}$), because the density--and therefore, the index of refraction which defines the Cerenkov angle--is low at the top of the atmosphere. Comparatively, light emitted in the extensive air shower will be emitted at broader angles with respect to the particle trajectory ($\sim0.2$--$2.0^{\circ}$). The DC light will also arrive $\sim$2 ns after the extensive air shower, because it has to traverse more material than the photons produced lower in the atmosphere. The simulation illustrated by Fig. \ref{fig:dc_timeangle} shows that by exploiting these two characteristics, the DC light can be identified.  The simulations were done using the Corsika Monte Carlo package, 6.2040 at an observation level of 1275 m above sea level \citep{1998FZKA6097}.

Cleanly separating the DC light from the EAS light is critical, because the DC light can be used to measure the charge of the incoming particle while the EAS light can be used to measure initial energy of the cosmic ray. The number of Cerenkov photons produced by an individual charged particle scales with the square of the charge of the particle \cite{0508-3443-6-7-301}.  Thus the measurement of the cosmic ray's electric charge depends only weakly on the hadronic interaction model used to simulate the events.  This technique was recently confirmed with the first ground-based detection DC light and the subsequent measurement of the energy spectrum of  five charge bands from protons to iron \citep{2007PhRvD..75d2004A}.

The DC signal is compact in both time and angular space, while the extensive air shower is more broad. The variation in arrival times and angles in the EAS light arise because the air shower light is emitted at a wide range of heights and from a number of secondary particles. A telescope that makes use of the timing and angular separation between the two signals will be able to measure the species of the primary particle to high energy, provided that it has enough effective area to account for the diminishing flux at higher energies.

 \begin{figure}[htbp] 
    \centering
 	\includegraphics[width=90mm]{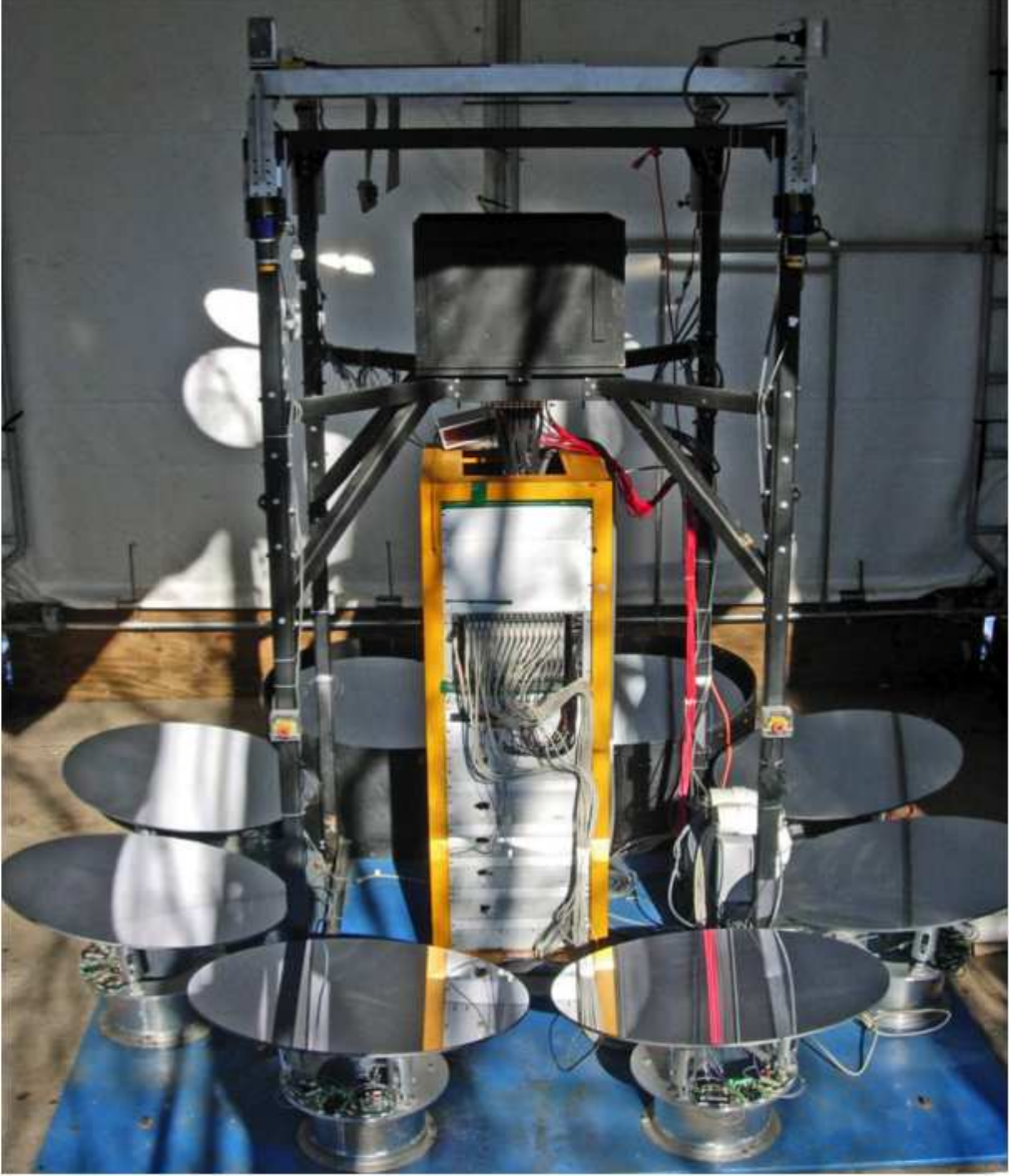}
   \caption[The TrICE Telescope]{The completed TrICE telescope with the baffle and electronics rack in the center of the telescope. Spherical mirrors with 4-m focal length line the base of the structure supporting the camera, Fresnel lens, and planar mirror. Each spherical mirror has a 1-m diameter. The height of the telescope, from the baseplate to the top of the planar mirror is 3 m.}
      \label{fig:trice_fulldet}
\end{figure}

The Track Imaging Cerenkov Experiment (TrICE), sited at Argonne National Laboratory and shown in Fig. \ref{fig:trice_fulldet}, serves as a test bed for new technologies that exploit the timing and angular characteristics of DC events. In particular, it employs a novel optical trigger and a camera composed of multi-anode photomultiplier tubes (MAPMTs) to image cosmic-ray events in both high- and low-resolution modes. The primary objectives were to verify the proposed technique via the detection of the Cerenkov light of air showers and to engineering a test facility for camera technology, while the secondary objective was to observe DC events.  

\begin{figure}[htbp]    \centering

   \subfloat[][Photons collected in an IACT-like Camera]{\label{fig:IACTcamera}\includegraphics[width=2.5in]{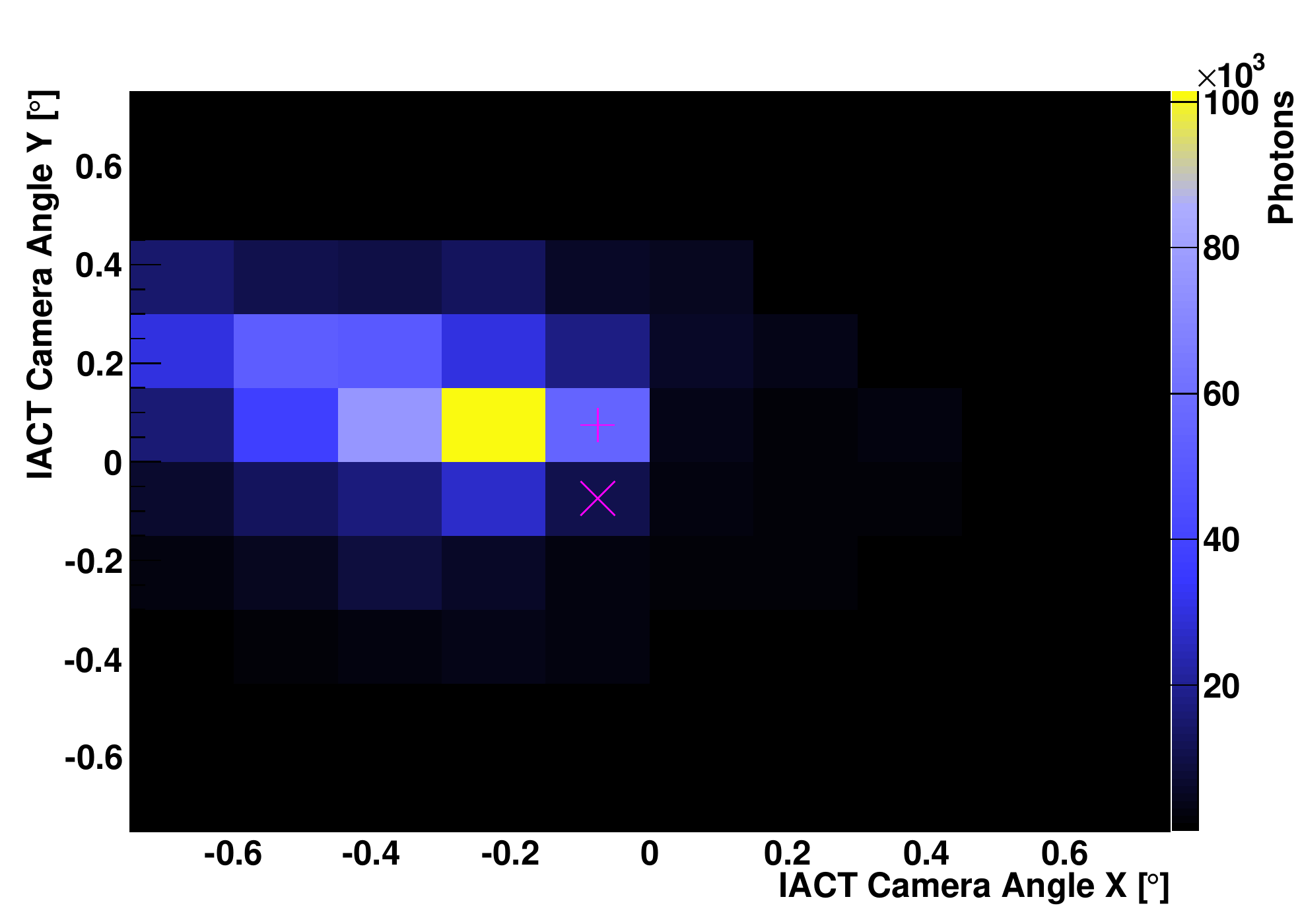}}\\
   \subfloat[][Photons collected in a TrICE-like Camera]{\label{fig:TrICEcamera}\includegraphics[width=2.5in]{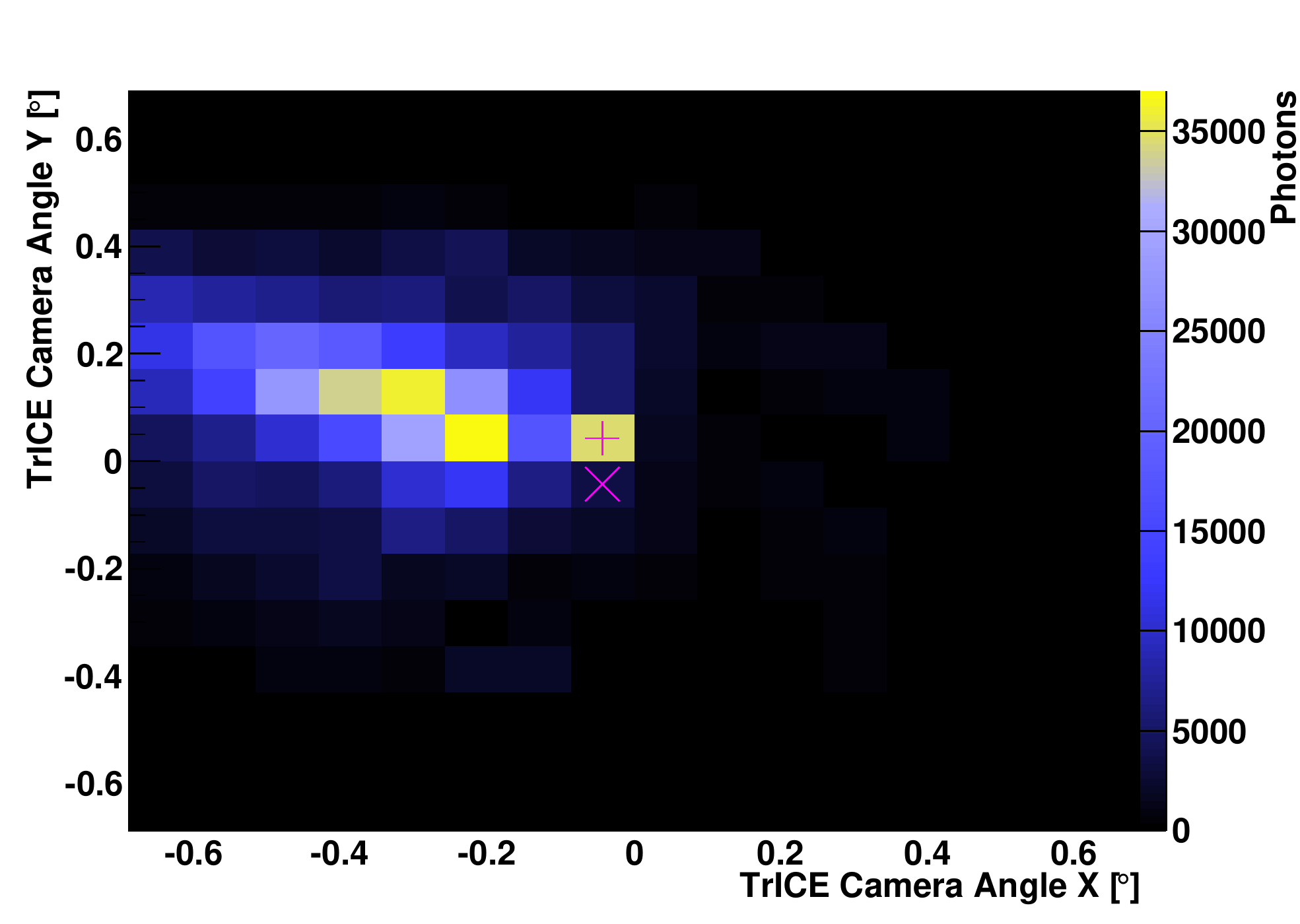}}
   \caption[Imaging DC light with TrICE and 2$^{nd}$-generation IACTs]{A comparison of the ability for a telescope with angular resolution comparable to a typical IACT camera (0.15$^{\circ}$, (a)) and a telescope with a TrICE-like camera (0.086$^{\circ}$, (b)) to separate DC light from EAS light from the simulation shown in Fig. \ref{fig:dc_timeangle}, landing 70 m from the telescope. The DC light in the front of the shower--indicated here by the $+$ and $\times$ symbols--is more prominent in the TrICE-like camera, because of the improved angular resolution. The pixel marked with the $+$ sign in (a) is composed of 57\% DC light; the pixel marked with a $\times$ sign, 10\%. Similarly, the pixel marked with the $+$ sign in (b) is composed of 89\% DC light; the photons in the pixel marked with a $\times$ sign, 33\%.}      \label{fig:em_ang_comp}

\end{figure}
Imaging air Cerenkov telescopes (IACTs) are designed to map the longitudinal development of an EAS into angular space \citep{2008RPPh...71i6901A}.  As demonstrated by the discovery of DC radiation by the H.E.S.S. collaboration, DC light can be detected by IACTs with 0.15$^{\circ}$  angular resolution \cite{2007PhRvD..75d2004A}. However, improving the angular resolution reduces the contamination of secondary air shower photons into the pixels containing mostly DC light, and therefore, reduces the dependence of the charge measurement on energy. To compare the power of the TrICE camera to a standard IACT, consider a  vertically-incident iron nucleus of energy 50 TeV that interacts 25 km above the observation level (Fig. \ref{fig:em_ang_comp}). When the IACT images it, the DC signal is easily visible in the camera plane and concentrated into one phototube, towards the front of the shower. However, the timing separation ($\sim$2 ns) can be ambiguous in the case of non-isochronous mirrors, and while the DC signal is visible, it is generally contaminated by air shower light because the angular separation between the DC light and the end of the shower, 0.2$^{\circ}$, is only slightly larger than the 0.15$^{\circ}$ viewing angle of a standard IACT photomultiplier tube \citep{2008ICRC....2..417W}. The TrICE telescope, with its improved angular resolution of 0.086$^{\circ}$, however, would image the DC light in multiple pixels and enhance the time separation between the two signals using the optics shown in Fig. \ref{fig:trice_optics_demo}.

\section{Description of the telescope}

\subsection{Optical Design}
\label{sec:trice_optdesign}
\begin{figure}[htbp] %
    \centering
   \includegraphics[width=90mm]{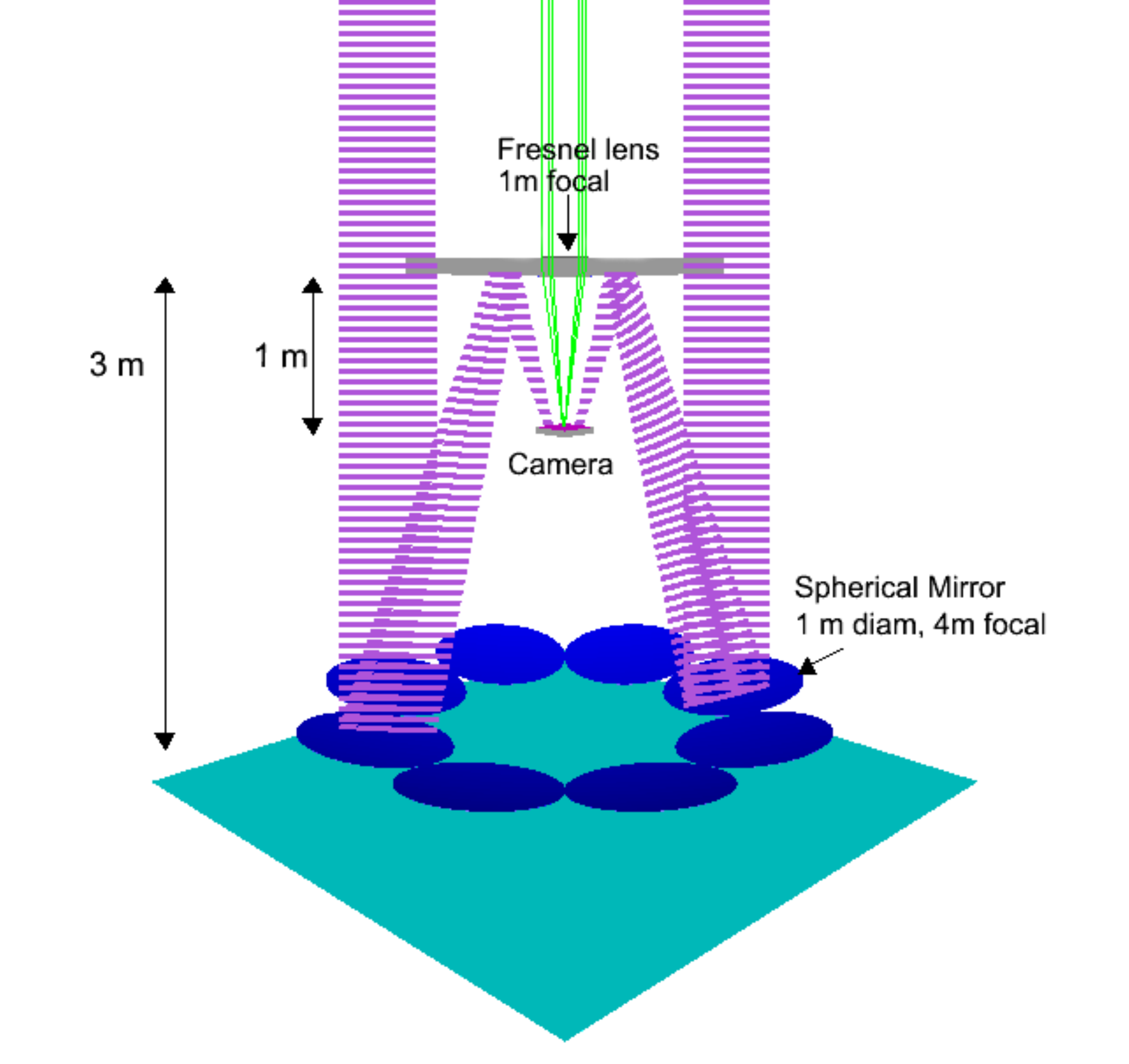} 
   \caption[The TrICE optical system]{A ray-traced simulation of TrICE. Collimated light impacting the spherical mirrors, shown in the dashed lines, is reflected onto the camera plane by the planar mirror, while the light impacting the Fresnel lens, shown in the solid lines, is focused directly onto the camera plane.}
   \label{fig:trice_optics_demo}
\end{figure}

TrICE employs a specialized optical design (Fig. \ref{fig:trice_optics_demo}), to provide a trigger through one light path and high-resolution imaging through a second. A Fresnel lens mounted directly above the camera plane enables a trigger signal that precedes a delayed and magnified image of the shower. Eight spherical mirrors are arranged on a square perimeter around the base of the telescope. They focus light, generating the high-resolution image onto the camera via a secondary planar mirror that also serves as a frame for a Fresnel lens.

TrICE features a primary mirror area of 6.4 m$^{2}$ and a 1.5$^{\circ}$ field-of-view, with eight 1-m spherical mirrors of focal lengths of 4 m. Each spherical mirror is mounted on 50.8 cm high pedestals. Lab tests determined the radii of curvature of all of the spherical mirrors to be within 0.64 cm of 7.945 m. The 3.7$^{\circ}$ wide field-of-view of the Fresnel lens collects light from a wider angle than the mirrors and the difference in magnification between these two systems is a factor of $\sim$4. The light from the air shower that is imaged by the coarse-grained optics of the Fresnel lens triggers the system.  The light from the mirrors, which comes from both the air shower and the primary particle, arrives nanoseconds later, because of the longer optical path length inherent to the mirror system. Using the Fresnel lens as an early trigger allows TrICE to first get a wide-field view of the air shower and later get a focused image of the shower and the DC light.

Four flat downward-facing mirrors were mounted in the grid using a silicone adhesive to make the planar mirror surrounding the Fresnel lens. In order to ensure the flatness of the planar mirror a series of measurements were made. First, micrometer measurements were taken around the edges of the mirrors. Next, a three-dimensional map of the mirrored surface was created using a coordinate-measuring machine with retractable arm which recorded the three-dimensional spatial coordinates of the mirror with respect to a well-defined origin. Overall, the lowest point recorded was 0.208 mm deviant from the average value. The flatness was confirmed to within a tolerance  of 0.02$^{\circ}$.

Typically air Cerenkov telescopes, which have pointing capabilities, use stars to align the mirrors with an object placed at infinity.  However, few stars of the necessary magnitude ($\sim$2 per year due to the ambient light) pass over the fixed pointing of TrICE, which means that the mirrors must be aligned relative to each other. A custom motor-control system was designed to ensure that alignment could be done rapidly. The alignment system allowed all of the spherical mirrors to be controlled simultaneously.  Commands given through a graphical user interface are sent through TTL lines to a mirror control module which drives the two actuators on each mirror. Hall effect limit switches prevent over-stressing the mirrors.  The planar mirror can be driven vertically to change the focal length of both optical paths. 

Relative alignment of the spherical mirrors was achieved by fixing a white LED, of diameter 0.56 cm, $\sim$11 m above the ground, such that the light would be focused $\sim$40 cm above the ground. The LED was first aligned with a laser placed in the center of the telescope on the ground and next with the light collected by the Fresnel lens. Then using a Starlight Xpress SXV-M7 CCD camera to record and analyze the point-spread function (PSF) at the height of the focal point, the images from the mirrors were aligned to one another.  This process was performed iteratively, because the focal plane shares space with the data acquisition system, which needed to be moved upon each iteration. Thus, four mirrors were aligned simultaneously, repeating the procedure until all combinations of mirrors were aligned to each other. The intensity of light at the focal plane had a 90\% enclosure region of diameter 0.6 cm. Deconvolution of the intensity function with the inherent width of the LED suggests a point-spread function of 2.2 mm which is smaller than the MAPMT camera pixel width of 6 mm.

\subsection{MAPMT Camera}
\label{sec:trice_mapmt_dacq}
The TrICE camera consists of an array of 16 MAPMTs, 256 pixels total.  Each pixel has an 0.086$^{\circ}$ field of view, in diameter and the entire camera has a fill factor of 0.79.  The tubes are mounted on a circuit board that in turn mounts onto the imaging plane of the telescope.  The camera incorporates a baffle to provide a shield from horizon light.  Additional functionality on the circuit board allows for separate monitoring of the MAPMT currents.  The analog signals are sent on short cables without pre-amplification to the electronics modules, where the integration of the current pulses and digitization of data are performed.

\begin{figure}[htbp]    \centering
   \includegraphics[width=90mm]{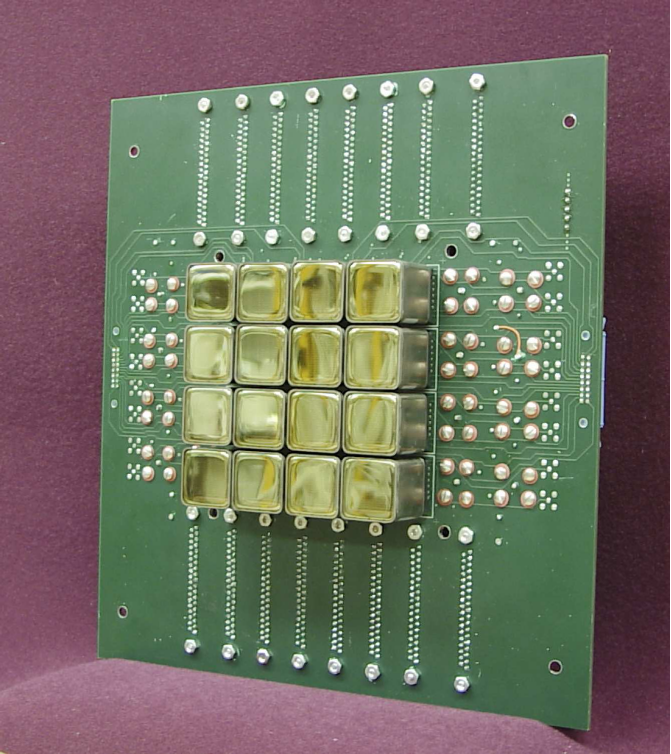} 
   \caption{The multi-anode photomultiplier tube camera installed on TrICE.}
   \label{fig:trice_mapmt}
\end{figure}

The TrICE camera consists of 16-channel Hamamatsu R8900 multi-anode photomultiplier tubes (MAPMTs) and is shown in Fig. \ref{fig:trice_mapmt} \citep{2008ICRC....2..469B}.
A number of requirements were placed on the photosensor including: that the sensor could distinguish single photoelectrons (PEs); that it have low crosstalk and the ability to handle pulsed operation in the presence of the large DC anode currents from the night sky background (NSB); that the sensor have good linear response over the dynamic range of interest (1--100 PEs); and that the sensor exhibit stable gain. Furthermore, the optics allowed for an optimum sensor pitch of approximately 5 mm. For details on the MAPMT selection see \cite{2008ICRC....2..469B}.

\begin{figure}[htbp] 
   \centering
   \includegraphics[width=90mm]{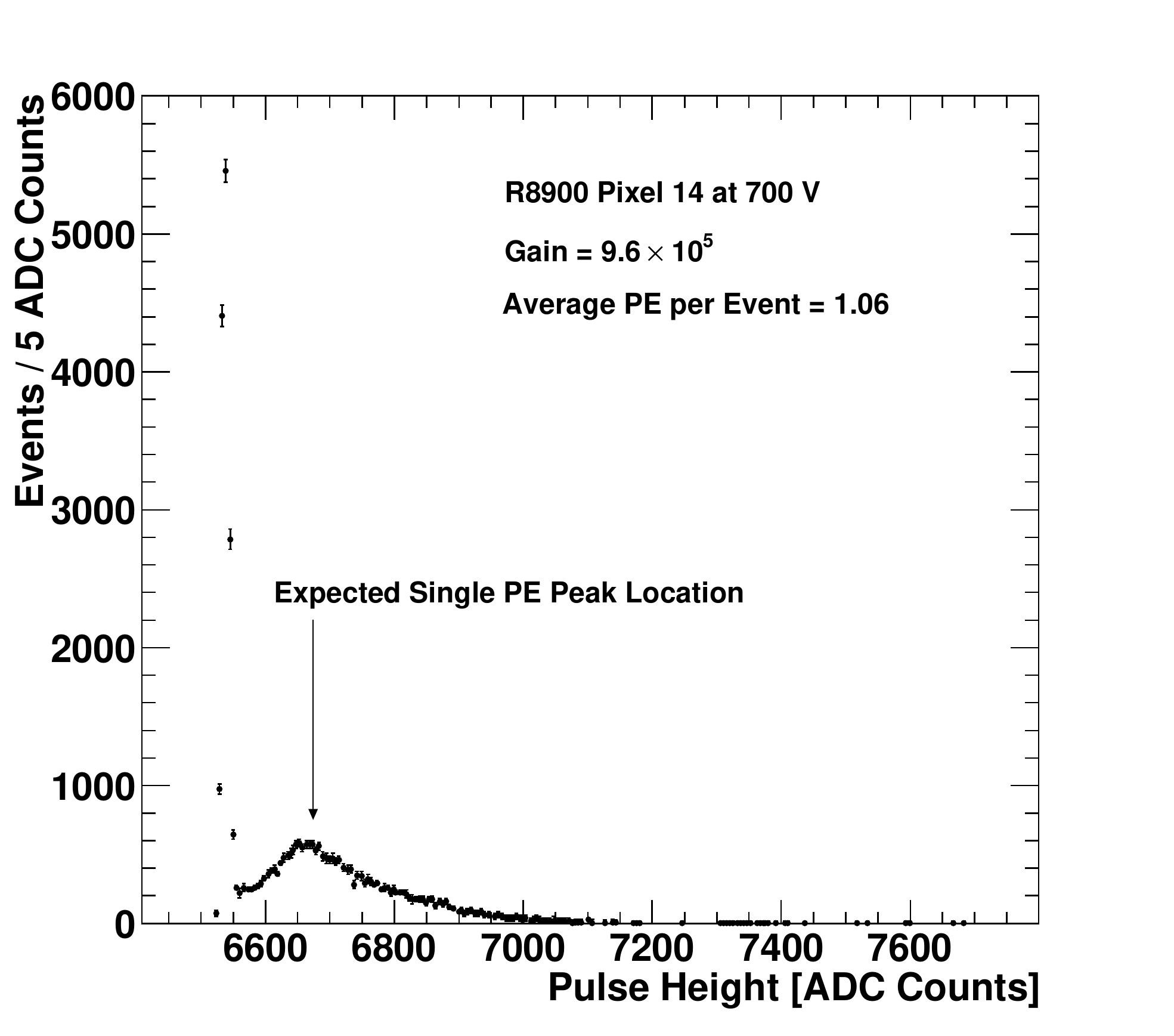} 
   \caption{The measurement of the single PE peak for one pixel in the R8900 MAPMT.}
   \label{fig:singlepe}
\end{figure}

Sensor characterization work was performed in a custom-designed dark box developed for specifically studying photosensors. The test stand included an LED, a reference PMT, an x-y stager with an automated fiber positioner, and an exponentially-graded neutral-density filter connected to a stepper motor and DAQ readout. The electronic read out consisted of a 16-bit ADC and an amplifier board to yield 1.144fC/ADC charge resolution. In this test stand, the R8900 exhibited a single PE peak (shown in Fig. \ref{fig:singlepe}), thus permitting the tracking and understanding the gain of the device.

Low crosstalk among adjacent pixels and good linear response affect the camera's ability to accurately characterize shower distributions. The R8900 MAPMTs showed low crosstalk in adjacent pixels, on the order of 2-3\%, and good linear response over the dynamic range of 1 to several 100 PEs. A deviation of ~5\% from linearity was observed at the highest levels, above approximately 200 to 500 PE \citep{2008ICRC....2..469B}. 

\begin{figure}[htbp] 
   \centering
     \subfloat[][Month-to-month relative gain stability as determined from a diffuse, uniform light source]{\label{fig:trice_relgains} \includegraphics[width=90mm]{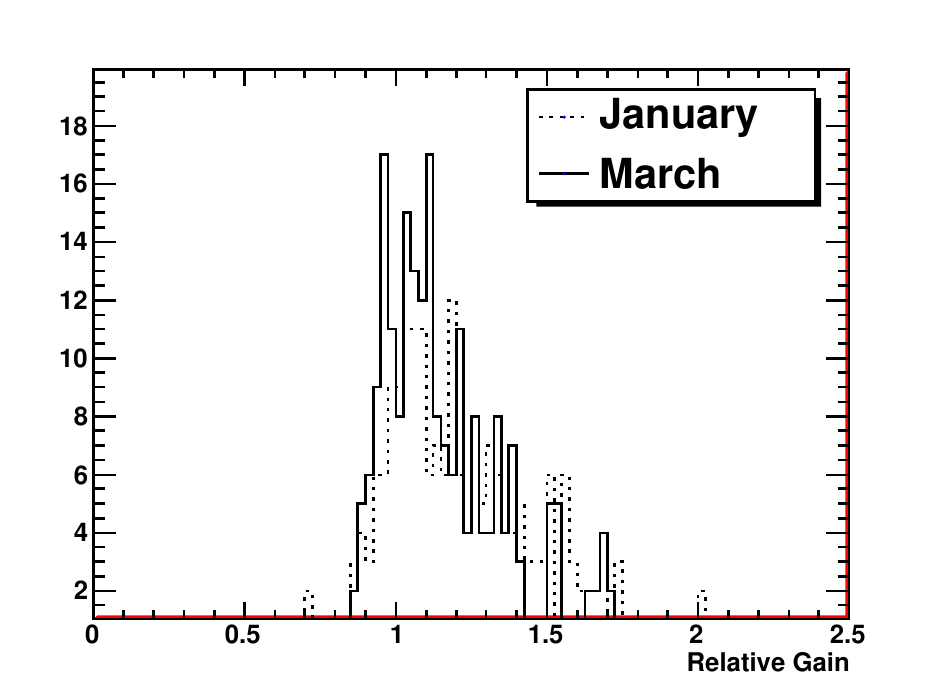} }\\
   \subfloat[][Map of relative pixel gains in the MAPMT camera as measured in the custom-built dark box.]{\label{fig:gainmap}\includegraphics[width=90mm]{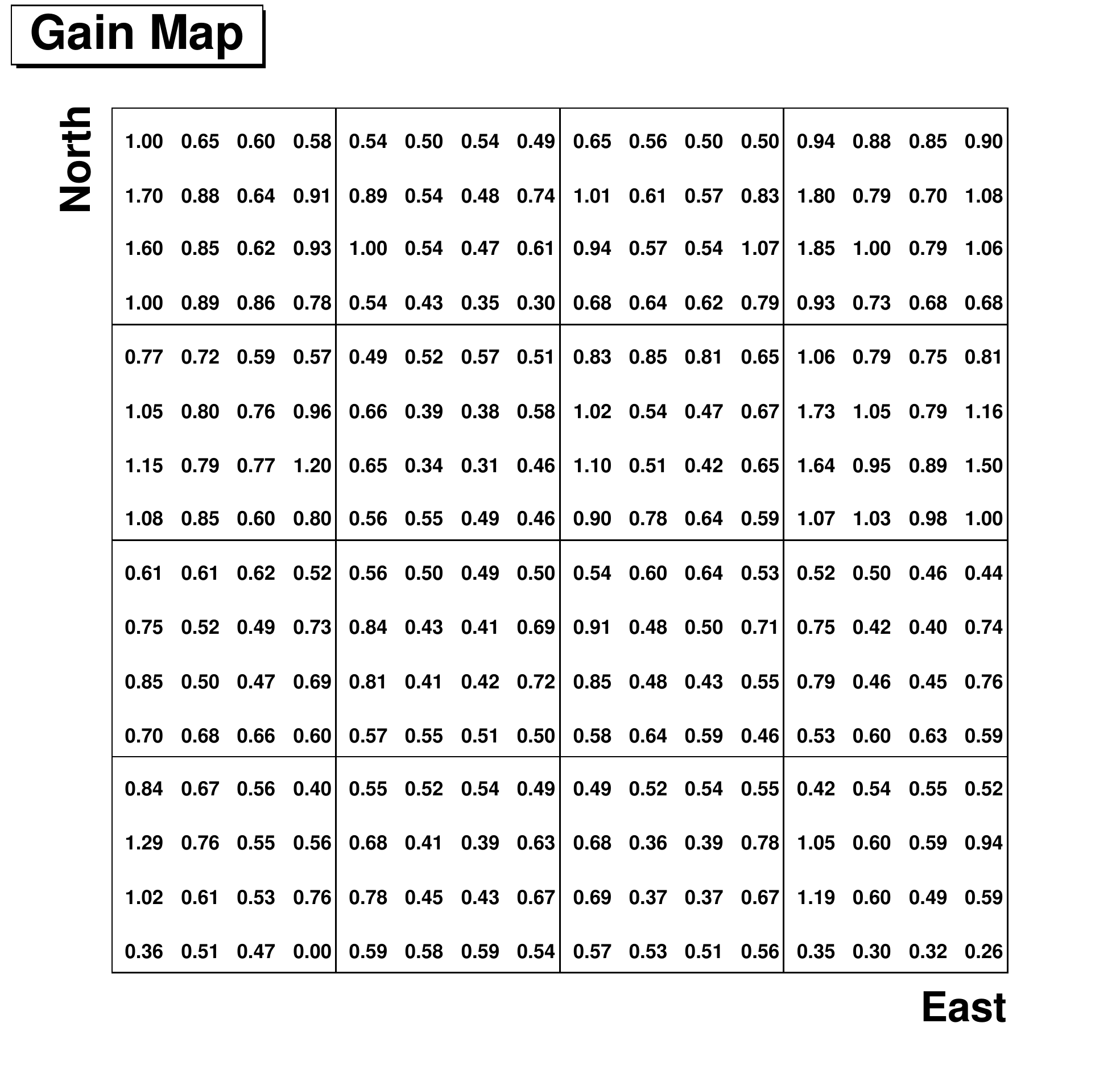} }
   \caption[Gain stability in TrICE]{Relative channel gains as determined by flat-fielding the camera over the course of a season in 2007 (a) and in lab tests (b).}
\end{figure}

Lab tests performed show that the R8900 pixel-to-pixel gain varies by a factor of 2 to 3 within an MAPMT, as shown in Fig. \ref{fig:gainmap}. Regular flat-fielding tests were performed to ensure that the camera maintained stable gains throughout the observing season. Using a diffuse light source mounted above the camera, the camera's response to a continuous light source can be monitored. Fig. \ref{fig:trice_relgains} shows that the distributions of individual gains are consistent from month to month.  Such gain measurements were conducted on a nightly basis to monitor the response over time.  Measurements made at different times throughout a single night show that intra-night variation is dominated by the warm-up effects during the first 15 minutes after high voltage is applied, and to a much lesser extent by temperature variations. Following the warm-up period, the gain response of the MAPMTs is largely stable. 

A measurement of the ambient light from the night-sky was made for the TrICE site. Tests using a baffle constructed around a single MAPMT showed that the anode current on the phototube without Cerenkov light (i.e., the night sky background rate NSB) was 1 $\mu$A per pixel, which required modifications made to the base of the MAPMTs enabled operation at reduced rates.

\subsection{Data acquisition}
Electronics to read out and digitize the signal from the MAPMTs were designed for the Near Detector of the MINOS Experiment \citep{2006ITNS...53.1347C}. The front-end electronics employ charge-integrating encoders, or QIEs, to perform the analog processing. The system can be used in two triggering modes: one in which data is read out as after a 10 $\mu$s ``fast spill'' and marked as such, and another in which the dynode signal from photomultiplier tubes is used as the trigger.  The TrICE telescope uses the latter triggering mode on the QIEs.

Two VME crates hold two types of modules: one designed for the digitization of MAPMT signals and one for the buffering of data record the signals from the PMTs. Analog signals from a 16-channel MAPMT travel to a corresponding 16-channel module housed in the front-end VME crate.  The QIEs continuously digitize at 53 mega samples per second to measure the integrated charge in 18.8 ns time slices. Upon receiving a trigger, the digitized data for 8 buffered time slices (150.4 ns total event window) are transferred to the separate back-end VME system.

The dynode signals from the 4 central MAPMTs of the camera are used to trigger the front-end electronics. A trigger is generated by the coincidence of at least two MAPMT dynode signals above a programmable threshold. In addition to the dynode-based trigger, the front-end electronics also accepts an external trigger signal. This mode is run separately from normal observations and is primarily used for measuring the effects of the night sky photon background on integrated charge values. 

The back-end VME data acquisition consists of 9U VME modules designed to receive and buffer the digitized signals until they can be transferred to disk. A GE Fanuc VMIVME 7766 single board computer running Linux controls the programming of the front-end and back-end VME modules and also handles the transfer and storage of buffered data. The back-end modules employ two memory buffers that switch on 20 ms interrupts. The VME CPU transfers data from the inactive buffer to disk while the active buffer fills with incoming data from the front-end modules. To keep pace with the 20 ms cadence of the buffer swap, the raw data are simply stored on disk and then later translated into time and channel ordered events as part of an additional event processing stage that can be run externally to the DAQ processing unit. 

\section{ High-Resolution Cosmic-Ray Air Showers }
\label{section:trice_hiresshowers}
Observations were made using the calibrated TrICE telescope from January through June, 2007. Operating thresholds were adjusted nightly to ensure rates between 40 and 120 Hz depending on the weather and the number of mirrors included in the observations. Observations where only the Fresnel lens or one spherical mirror is uncovered were used as diagnostic tools during regular observations.

\begin{figure}[htbp] 
   \centering
   \subfloat[][]{\label{fig:trice_event_chageint} \includegraphics[height=45mm]{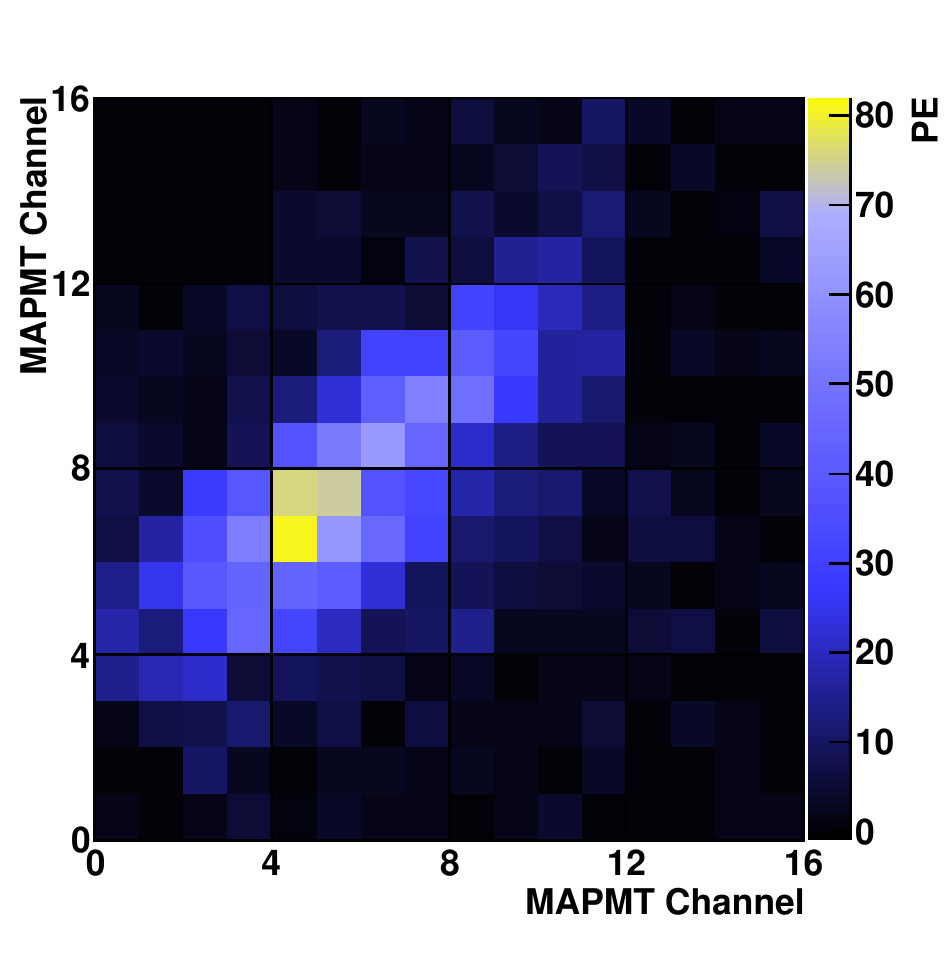}}
   \subfloat[][]{\label{fig:trice_event_deg} \includegraphics[height=45mm]{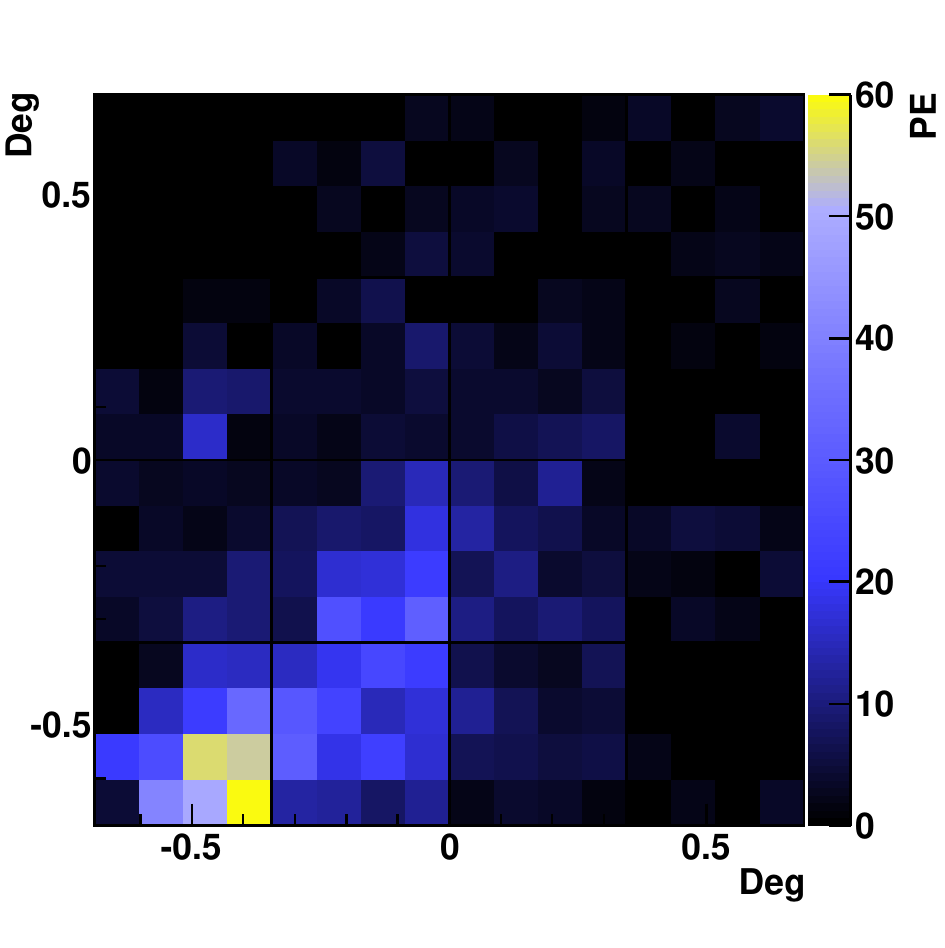}}
   \caption[High-resolution air shower images from TrICE]{Two high-resolution images of different cosmic-ray air showers taken with the TrICE telescope, plotted in channel numbers in (a) and the camera angle in (b).  Each pixel represents $0.086^{\circ}$ in camera angle.  The number of photoelectrons in each pixel is shown, integrated over the first three time slices in the events.}
      \label{fig:trice_hires_crshowers}
\end{figure}

The images of extensive air showers are among the highest resolution images ever recorded. Fig. \ref{fig:trice_hires_crshowers} shows two such events taken showing an extended shower with a peak slightly displaced from the center of the shower.  Substructure in the air showers events is also visible. Note that the structure is smaller than the scale of one MAPMT, indicating that the camera is capable of measuring intensity differences for each pixel.

Almost all of the signal in the TrICE data is contained in the first three time-slices, while the last time-slice contains the pedestal information.  The first three time-slices for each pixel are summed to give the time-integrated image of each event. While the TrICE optical system was designed to exploit the few nanosecond time separation between the DC light and the EAS light, the electronics available were not fast enough. In order to build a detector that uses the time separation to identify DC events, at least 1 giga sample per second sampling is required \cite{2008ICRC....2..417W}.

The electronic pedestal remains fairly constant at a value between 100 and 120 digital counts and is subtracted from every event. After flat-fielding with the lab-measured gain values for each MAPMT, the images are cleaned using a fixed threshold cut set at 5 photoelectrons for pixels within the image and 2 photoelectrons for pixels on the edges of the image.   
 
The images are then fit to an ellipse using a function that finds the borders of an ellipse. A copy of the image -- hereafter known as the ellipse map -- sets image pixels to a constant value and the border pixels to a value between zero and one . If the point being tested in the fit is inside the ellipse, then the function returns a value of 1 and if it is outside the ellipse, then it returns a value of 0.  On the edges, the fit function returns a value between zero and one to ensure that the fit converges. The final fit gives information about the size, shape and orientation of the image in the camera plane. 

\begin{figure}[htbp]
	\centering
		\includegraphics[width=90mm]{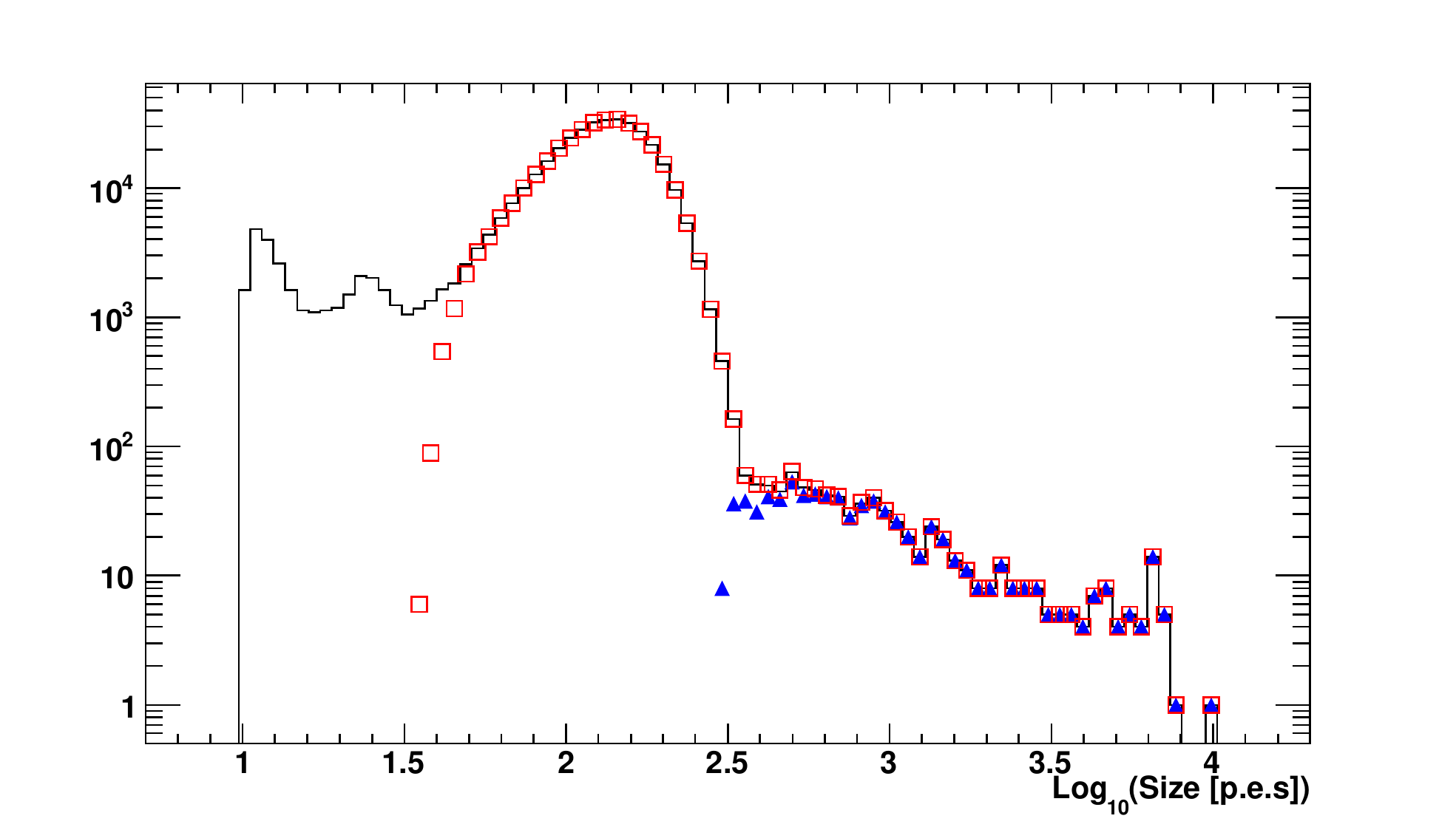}
		\caption{The distribution of event shower size in photoelectrons for three nights of data taken under similar conditions (histogram). A selection requiring 6 pixels passing image and border thresholds described in the text (squares) separates well-defined cosmic-ray showers from background events and noise seen in the two peaks at low sizes.  Showers that trigger at least 3 PMTs and at least 48 channels are shown in the triangles.}
	\label{fig:size}
\end{figure} 

The data shown in Fig. \ref{fig:size} represent the data taken during favorable weather conditions and comprise 8.5 hours of live time. The Cerenkov light in pixels contributing to the image is summed to calculate the total shower size for each event.  As shown in Fig. \ref{fig:size}, high-energy showers are selected from the background by requiring that at least six neighboring pixels pass the fixed image and border thresholds. An analysis threshold requires that  $\sim$32 photoelectrons contribute to a coherent shower image. For comparison, the simulation shown in Fig. \ref{fig:em_ang_comp} has a size of 17000 photoelectrons, a number which reduces to 16000 when you remove the contribution from the direct Cerenkov light. This calculation assume an average mirror reflectivity of 70\%, a single photoelectron ratio of 1.06 and MAPMT quantum efficiency of 35\%. On average, showers that include at least 6 channels in their image have a size of 145 p.e.s, a factor of 100 lower than the size of the simulation. The event shown in Figure \ref{fig:trice_event_chageint} has a size of 2208 p.e.s; the one in Figure \ref{fig:trice_event_deg}, 1166 p.e.s. Such showers fall meet the requirement of triggering at least 48 channels and at least 3 PMTs, the distribution of which are indicated by the triangles in Figure \ref{fig:size}. Therefore likely have an energy that is a factor of 10 lower than the simulation shown in Figures \ref{fig:dc_timeangle} and \ref{fig:em_ang_comp}.

\begin{figure}[htbp]
	\centering
		\includegraphics[width=90mm]{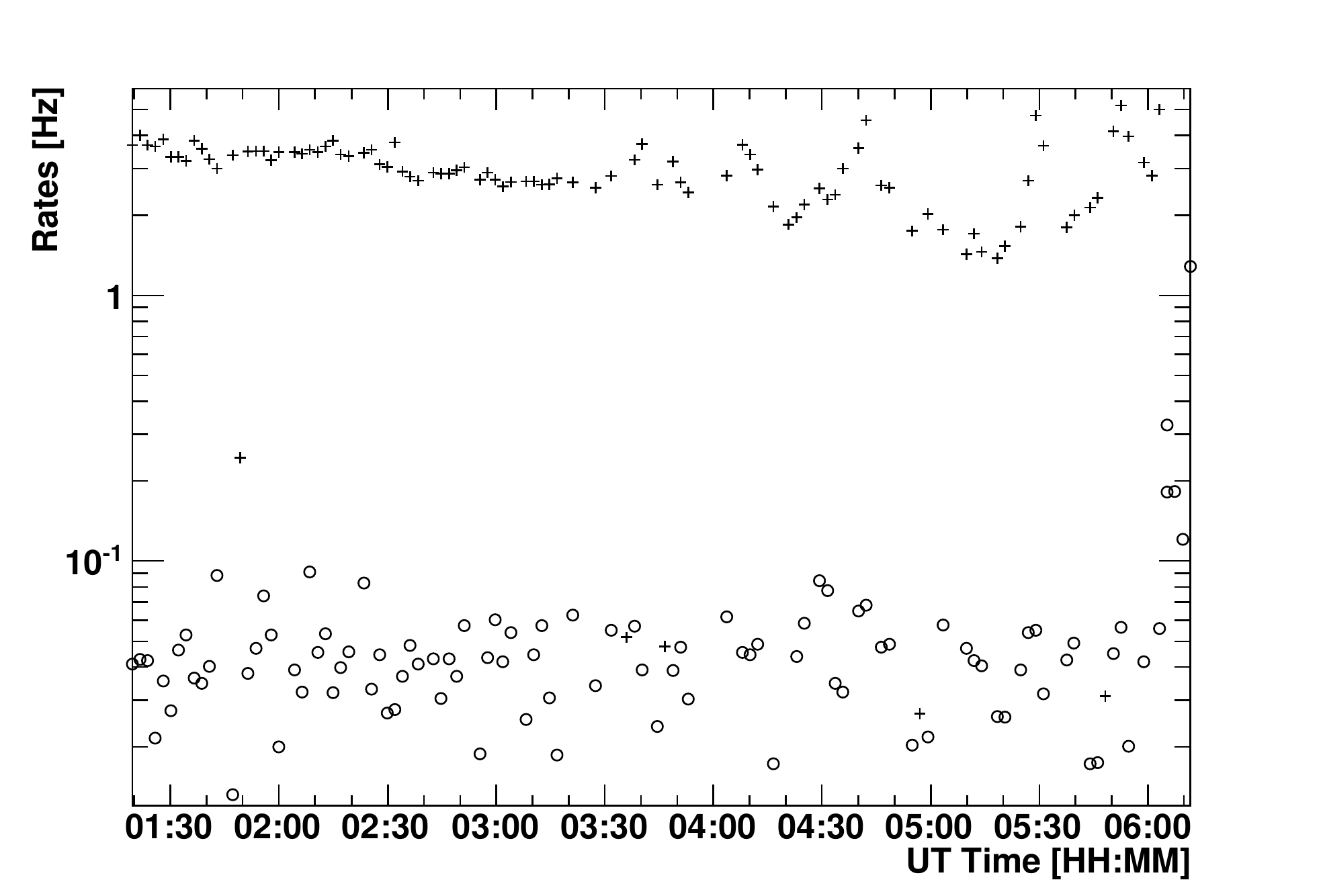}
		\caption{The trigger rates in which one pixel meets the image threshold requirement (crosses) and cosmic-ray event rates (circles) are shown for one night of observation.  Cosmic-ray events are selected by requiring that at least 6 pixels contribute to the image after cleaning.}
	\label{fig:rates}
\end{figure}

Typical trigger rates depend on the night-sky conditions, but range between 40 Hz and 120 Hz.  After requiring that a minimum of six pixels are included in an event, the event trigger rates are reduced to 0.1 Hz. The two rates for a single night of observations are shown in Fig. \ref{fig:rates}. For this particular night, the rates increase towards the end of data taking as the weather started to worsen. On a regular basis, the telescope was operated with only a single mirror was exposed to check the alignment and optical uniformity among the spherical mirrors. Rates during these single mirror runs typically ranged between 1 and 2 Hz. 

\section{Conclusions}
\label{sec:trice_outlook}
The main successes of the TrICE telescope include the stable operation of a MAPMT camera in a telescope employing the imaging atmospheric Cerenkov technique and the demonstration of a novel optical system, including a proof-of-concept automated mirror alignment system. The technologies tested during the construction and operation of this telescope may be considered for use in the design of an observatory dedicated to cosmic-ray observations.  

The optical system provided both an optical trigger and the convenience of a fixed-mount design. Alignment of the fixed-mount optical system proves to be difficult when the telescope must be operated in weather conditions, but using a remotely-controlled alignment system and a fixed point-source, relative alignment of the mirrors has been shown to produce a point-spread function smaller than the MAPMT pixel size.  Note that because of the improved angular resolution, the optical quality must be better than what is found in typical IACTs.

The TrICE telescope demonstrates that a MAPMT camera can be stably operated in a cosmic-ray telescope under high ambient light conditions with little variation in operating conditions.  Well-resolved images of the cosmic ray air showers were seen in the TrICE data, demonstrating the level of precision available to the next generation of $\gamma$-ray and cosmic-ray telescopes that will use cameras with comparable angular resolution. 

Although no DC events were detected in TrICE, this is consistent with the limited $\sim200$ m$^{2}$ s sr exposure which was obtained with the instrument. Upon the conclusion of the 2007 observing season, efforts were shifted from trying to identify DC events with TrICE to using the DC technique to measure the iron spectrum in VERITAS data \citep{2010PhDT........37W}. Nevertheless, the lessons learned from the operation of this instrument can help guide the design of a future DC instrument, be it a dedicated observatory \citep{2008ICRC....2..349W} or a the planned next-generation IACT array, the Cherenkov Telescope Array (CTA) \citep{2010arXiv1008.3703C}.

\section{Acknowledgements}
This research is supported by the U.S. Department of Energy, the U.S. National Science Foundation, and the Kavli Foundation.  We acknowledge the technical contributions of Richard Northrop, Gary Kelderhouse and Bob Metz.

\bibliographystyle{model1a-num-names}
\bibliography{TrICE}
\end{document}